# Analytical Green's Function of Multidimensional Boltzmann Transport Equation for Modeling Hydrodynamic Second Sound

Xin Qian[1]*, Chuang Zhang[2], Te-Huan Liu[1]*, and Ronggui Yang[1,3]*

[1] School of Energy and Power Engineering, Huazhong University of Science and Technology, Wuhan 430074, China.

[2] Department of Physics, Hangzhou Dianzi University, Hangzhou 310018, China.

[3] College of Engineering, Peking University, Beijing 100871, China.

Corresponding Emails: xinqian21@hust.edu.cn, thliu@hust.edu.cn, ronggui@pku.edu.cn

**ABSTRACT**

Hydrodynamic second sound can be generated by heat pulses when the phonon-phonon interaction is dominantly momentum conserving, and the propagation of the temperature field becomes wavelike rather than diffusive. While the Boltzmann transport equation (BTE) has been widely applied to study phonon dynamics and thermal transport at the nanoscale, modeling the hydrodynamic transport regime remains challenging. The widely used relaxation time approximation (RTA) treats all phonon interactions as resistive without considering momentum conservation, resulting in the absence of phonon hydrodynamics. Rigorously solving BTE by inverting the full scattering matrix, however, is extremely computationally demanding and has been only applied to model one-dimensional temperature variations. Here, we present an analytical Green's function formalism for solving multidimensional Boltzmann transport equation (BTE) using phonon properties from first-principles calculations. This formalism involves Callaway's scattering approximation with separate relaxation times for momentum-conserving and momentum-destroying scattering events. The Green's function captures phonon dynamics in a wide range of temperature, spatial, and temporal scales, and successfully reproduces the transition from ballistic, hydrodynamic, to diffusive transport regimes. Our method avoids the computationally demanding inversion of large scattering matrices and shows good accuracies in predicting the temperature oscillation in ultrafast pump-probe characterizations with different geometries of thermal excitation.

# I. INTRODUCTION

Thermal transport in semiconductors or dielectrics is typically governed by the random walk of phonons[1]. Such diffusive transport regime breaks down at small lengths or time scales when phonons transport ballistically without experiencing enough scattering to establish a local equilibrium [2,3]. Ballistic transport of phonons has been studied extensively [4-9] due to the significance in engineering thermal conductivity of functional materials, and thermal management of nano/microelectronic devices. In contrast to ballistic transport, hydrodynamic phonon transport emerges when phonon-phonon interaction is frequent and mostly momentum-conserving [10,11]. In the hydrodynamic regime, the propagation of the temperature field becomes wavelike under pulsed heating, which is known as the second sound [12]. The second sound was first observed in superfluid He below 2 K [13], and later in solids such as Bi at 1.2 K ~ 4 K [14] and NaF in 10 ~ 18 K [15]. Such "wavelike" transport of heat energy can serve as an efficient vehicle for removing heat from hot spots, but it was only observed at cryogenic temperatures for nearly 50 years. It was not until the recent few years that research interests in second sound was renewed due to the prediction and observation of second sound in graphitic materials close to room temperature (100 ~ 200 K) [16-18], and the discovery of wavelike signatures in thermal phase lags under rapidly varying temperature field in Ge at room temperature [19]. Specifically for transient thermal grating (TTG)[16] and transient thermoreflectance[20] measurements, the second sound will result in an abnormal cooling occurring within a few nanoseconds after the heat pulse is imposed on the sample, because of the phase inversion caused by the constructive interference of the hydrodynamic temperature waves.

Theoretically, the second sound can be modeled phenomenologically using a damped wave equation (DWE) [21,22], manybody correlation functions [23], or the Boltzmann transport equation (BTE) [18,24]. While the DWE provides a convenient way to study the wavy transport behaviors and to process photothermal measurement signals for identifying second sound, it lacks the power to predict the temperatures and lengthscales for realizing wavelike thermal transport, as the coherence time of second sound ($\tau_{ss}$) and the propagation length is treated as empirical parameters to be fitted from experimental data [21, 22]. On the other hand, the correlation function formulation developed by Enz [23] was underpinned by studying the evolution of local equilibrium density matrices, from which the dispersion relation and frequency window for observing second sound can be derived without empirical parameters. However, direct comparisons between the local correlation function theory with experimental results are usually difficult, due to the experimental challenges in directly probing spatial or temporal correlation functions in temperature or phonon momentum. The theoretical framework of BTE has been widely used for studying phonon transport, including first-principles prediction of thermal conductivity[25, 26], modeling temperature and heat dissipation in nanodevices [9, 27], and in studying hydrodynamic phonon transport [24]. The scattering term in BTE is either treated under relaxation time approximation (RTA) or fully calculated by first-principles anharmonic lattice dynamics. RTA assumes all phonon-phonon scattering events as resistive. It will not only underestimate the thermal conductivity of materials with strong N-scattering such as graphite and diamond, but also miss the entire hydrodynamic phonon transport regime due to the lack of momentum conservation for N-scattering processes. On the other hand, solving BTE with the full scattering matrix can effectively capture the collective phonon dynamics as the momentum conservation is automatically satisfied [28]. This has been done by iterative solution in thermal conductivity calculations, while most of the phonon transport modeling involving localized hot spots and interfaces has been performed under RTA[27, 29]. Recently, Chiloyan *et al.* [30] derived the exact Green's function of BTE by inverting the full scattering matrix and modeled second-sound meditated temperature oscillations in TTG experiments. Inverting the full scattering matrix is extremely expensive, especially with a dense mesh in the Brillouin zone, such that only one-dimensional problems are computationally feasible. Compared with RTA, Callaway's scattering approximation (CSA) with two distinct relaxation times for N- and R-scattering events represents a better approximation to the full scattering matrix, and it has been widely used in analyzing steady-state hydrodynamic transport, including the Knudsen minimum [31] and the phonon Poiseuille flow [18]. While the direct numerical solution of BTE under CSA has been performed for both steady-state and transient phonon transport in various geometries [32-35], gray model or empirical phonon lifetime relations are typically used for phonon dispersions and relaxation times, with few studies on steady-state hydrodynamic transport have incorporated *ab initio* phonon properties in the entire Brillouin zone [36].

In this work, we derived the analytical expression for Green's function of multi-dimensional BTE under CSA (hereafter abbreviated as GF-CSA), which can be used to predict transient temperature responses to arbitrary shape of heat sources in infinite or semi-infinite domains, using fully first-principle phonon dispersions and lifetimes. Compared with the GF calculated by inverting the full scattering matrix, the separation of N and R processes in CSA enabled analyzing the asymptotic behavior and phonon dynamics at different scattering limits. We analytically showed that: (i) the damped wave equation emerges from the GF-CSA formalism when N-scattering is dominating, (ii) the GF-CSA reduces asymptotically to RTA when R-scattering is dominating, and (iii) the ballistic and Fourier transport regime is restored at the limits $\Omega\tau \gg 1$ and $\Omega\tau \ll 1$ respectively, with $\Omega$ the heating frequency and $\tau$ the phonon relaxation time. Finally, GF-CSA successfully captures the temperature oscillations in TTG and transient thermoreflectance with a ring-shaped heat source (ring-TTR), showing excellent agreement with experimental observations. Our work provides a simple formalism for modeling multi-dimensional and transient hydrodynamic phonon transport.

## II. THEORETICAL DERIVATION OF GREEN'S FUNCTION

To obtain the Green's function of hydrodynamic phonon transport, we consider solving the BTE under a point heat pulse:

$$\frac{\partial n_\mu}{\partial t} + \boldsymbol{v}_\mu \cdot \nabla n_\mu = \frac{n_\mu^0(T_R) - n_\mu}{\tau_\mu^R} + \frac{n_\mu^d(T_N) - n_\mu}{\tau_\mu^N} + \frac{\dot{Q}_\mu}{\hbar \omega_\mu} \delta(\boldsymbol{r})\delta(t), \tag{1}$$

where $\delta(\boldsymbol{r})$ and $\delta(t)$ are Dirac delta functions in the spatial and time domains, the subscript $\mu = (\boldsymbol{q}, s)$ denotes the phonon mode with wavevector $\boldsymbol{q}$ at the $s$-th phonon branch, $n_\mu$ is the nonequilibrium phonon distribution, $\boldsymbol{v}_\mu = \nabla_{\boldsymbol{q}} \omega_\mu$ is the group velocity ($\omega_\mu$ is the angular frequency of phonon mode $\mu$), $\tau_\mu^N$ and $\tau_\mu^R$ are respectively the relaxation time for momentum-conserving N processes and momentum-reversing R-scattering events, including the Umklapp processes, isotope scattering, and boundary scattering. $\dot{Q}_\mu$ is the energy input rate into the phonon mode $\mu$. For a thermalizing heat source, $\dot{Q}_\mu$ is determined by the modal heat capacity $C_\mu$ as $\dot{Q}_\mu = \dot{Q}_0 C_\mu / C$, where $\dot{Q}_0$ is the total heat generation rate; $C_\mu = k_B (\hbar \omega_\mu / k_B T_0)^2 [n_\mu^0 (n_\mu^0 + 1)]$; and $C = \sum_\mu C_\mu$ is the total specific heat capacity. Since this work seeks a Green's function solution to the BTE, a point heat pulse with unit power is used here, and the total heating rate $\dot{Q}_0 = 1$ throughout this article. $\dot{Q}_\mu$ is therefore reduced to $\dot{Q}_\mu = C_\mu / C$, which is the fractional contribution of mode $\mu$ to total the specific heat $p_\mu = C_\mu / C$. Under CSA, the momentum-destroying and momentum-conserving scatterings are separately relaxed to a local equilibrium $n_\mu^0(T_R)$ and a drifted local equilibrium $n_\mu^d(T_N)$:

$$n_\mu^0(T_R) = \frac{1}{\exp\left(\frac{\hbar \omega_\mu}{k_B T_R(\boldsymbol{r})}\right) - 1} \tag{2}$$

$$n_\mu^d(T_N, \boldsymbol{u}) = \frac{1}{\exp\left[\frac{\hbar(\omega_\mu - \boldsymbol{q} \cdot \boldsymbol{u})}{k_B T_N(\boldsymbol{r})}\right] - 1} \tag{3}$$

where $T_R(\mathbf{r})$ and $T_N(\mathbf{r})$ are pseudo-temperatures for the local equilibrium of R-scattering and N-scattering events[35], respectively. Here, $\mathbf{q}$ is the phonon wavevector, and $\mathbf{u}$ is a collective drift velocity independent of phonon mode indices.

In the near-equilibrium transport regime, deviation in phonon populations can be linearized:

$$n_\mu - n_\mu^0(T_R) = n_\mu^1 - \frac{C_\mu \Delta T_R}{\hbar \omega_\mu}, \tag{4}$$

$$n_\mu - n_\mu^d(T_N) = n_\mu^1 - \left(\frac{C_\mu \Delta T_N}{\hbar \omega_\mu} + \frac{C_\mu T_0}{\hbar \omega_\mu^2} \mathbf{q} \cdot \mathbf{u}\right) \tag{5}$$

where $n_\mu^1 = n_\mu - n_\mu^0(T_0)$ is the deviation in phonon populations from thermal equilibrium at the temperature $T_0$. $\Delta T_{N,R} = T_{N,R} - T_0$ are pseudo-temperature changes quantifying the energy deviations from the equilibrium due to N and R scatterings respectively. Note that $\Delta T_N$ and $\Delta T_R$ are not measurable temperature rises, because one cannot distinguish detailed N- and R-scattering events when measuring a local temperature. The near-equilibrium assumption generally holds when modeling ultrafast pump-probe signals of second sounds. In pump-probe experiments, the temperature rise must be carefully controlled in the linear regime, such that the optical diffraction or the reflectance change scale linearly with the temperature rise $\Delta T$ [20]. However, when extremely high heat flux is imposed to drive the system beyond the linear regime, BTE must be solved nonperturbatively by considering temperature-dependent phonon properties, which is beyond the scope of this article.

By defining the energy deviation $g_\mu = \hbar \omega_\mu n_\mu^1$ and by substituting Eqs. (4-5) into Eq. (1), the BTE is simplified to:

$$\frac{\partial g_\mu}{\partial t} + \mathbf{v}_\mu \cdot \nabla g_\mu = \frac{C_\mu \Delta T_R - g_\mu}{\tau_\mu^R} + \frac{C_\mu \Delta T_N - g_\mu}{\tau_\mu^N} + \frac{C_\mu T_0}{\omega_\mu \tau_\mu^N} \mathbf{q} \cdot \mathbf{u} + p_\mu \delta(\mathbf{r})\delta(t) \tag{6}$$

We now consider solving this equation in a semi-infinite domain with a specular boundary condition at the surface. Under this assumption, the domain can be extended to infinity with even symmetry. We can then solve Eq. (6) by performing the Fourier transform as follows:

$$\hat{f}(\Omega, \boldsymbol{\xi}) = \iint f(t, \boldsymbol{r}) e^{-i\Omega t - i\boldsymbol{\xi} \cdot \boldsymbol{r}} dt d\boldsymbol{r} \qquad (7)$$

where $f$ and $\hat{f}$ are the real-domain and Fourier-domain representations of a generic function, and the integral domain is $(-\infty, +\infty)$ for all temporal and spatial coordinates. The Fourier-domain coordinates $(\Omega, \boldsymbol{\xi})$ are correspondingly the temporal and spatial frequencies, conjugate to the time and spatial coordinates $(t, \boldsymbol{r})$ in the real domain. Eq. (6) is transformed into an algebraic equation in the Fourier domain:

$$i(\Omega + \boldsymbol{v}_\mu \cdot \boldsymbol{\xi}) \hat{g}_\mu = \frac{C_\mu \Delta \hat{T}_R - \hat{g}_\mu}{\tau_\mu^R} + \frac{C_\mu \Delta \hat{T}_N - \hat{g}_\mu}{\tau_\mu^N} + \frac{C_\mu T_0}{\omega_\mu \tau_\mu^N} \boldsymbol{q} \cdot \boldsymbol{u} + p_\mu \qquad (8)$$

The energy deviation in the Fourier domain can be solved as:

$$\hat{g}_\mu = \chi_\mu \left[ C_\mu (\rho_\mu \Delta \hat{T}_R + \eta_\mu \Delta \hat{T}_N) + \eta_\mu C_\mu T_0 \frac{\boldsymbol{q} \cdot \boldsymbol{u}}{\omega_\mu} + p_\mu \tau_\mu \right], \qquad (9)$$

where $\tau_\mu$ is the total relaxation time defined as $1/\tau_\mu = 1/\tau_\mu^N + 1/\tau_\mu^R$, and $\rho_\mu = \tau_\mu / \tau_\mu^R$, $\eta_\mu = \tau_\mu / \tau_\mu^N$ is the relative strength of R-scattering and N-scattering events, and $\rho_\mu + \eta_\mu \equiv 1$. When $\eta_\mu \to 0$, the Callaway scattering model decays to simple RTA, and $\eta_\mu \to 1$ corresponds to the limit of all scatterings being momentum-conserving. $\chi_\mu$ is the linear susceptibility of phonon mode $\mu$, defined as:

$$\chi_\mu = \frac{1}{1 + i\Omega \tau_\mu + i \boldsymbol{F}_\mu \cdot \boldsymbol{\xi}} \qquad (10)$$

$\boldsymbol{F}_\mu = \boldsymbol{v}_\mu \tau_\mu$ is the mean free displacement (MFD). When R-processes are dominating, the lifetime $\tau_\mu$ can be evaluated using Fermi's golden rule at equilibrium[37]. For materials with nonnegligible N-processes, the MFD needs to be solved from [38]:

$$\frac{1}{N_q} \sum_{\lambda \nu} \left[ (\omega_\mu \boldsymbol{F}_\mu + \omega_\lambda \boldsymbol{F}_\lambda - \omega_\nu \boldsymbol{F}_\nu) \Gamma^+_{\mu\lambda\nu} + \frac{1}{2} (\omega_\mu \boldsymbol{F}_\mu - \omega_\lambda \boldsymbol{F}_\lambda - \omega_\nu \boldsymbol{F}_\nu) \Gamma^-_{\mu\lambda\nu} \right] \\ + \sum_\lambda \Gamma_{\mu\lambda} (\omega_\mu \boldsymbol{F}_\mu - \omega_\lambda \boldsymbol{F}_\lambda) = \omega_\mu \boldsymbol{v}_\mu, \qquad (11)$$

where $\mu, \lambda, \nu$ are phonon mode indices, $\omega$ denotes the phonon frequency, $N_q$ is the total number of mesh points in the Brillouin zone, $\Gamma^+_{\mu\lambda\nu}$ is the transition rate for the absorption three-phonon process $\mu + \lambda \to \nu$, $\Gamma^-_{\mu\lambda\nu}$ denotes the transition rate for the emission three-phonon process $\mu \to \lambda + \nu$, and $\Gamma_{\mu\lambda}$ denotes the elastic isotope scattering rates. The iterative solution of MFDs is performed using the ShengBTE package[25]. After obtaining the MFD, the effective mean free paths and relaxation times are calculated as:

$$\Lambda_\mu = \frac{\boldsymbol{v}_\mu \cdot \boldsymbol{F}_\mu}{|\boldsymbol{v}_\mu|}, \qquad \tau_\mu = \frac{\boldsymbol{v}_\mu \cdot \boldsymbol{F}_\mu}{\boldsymbol{v}_\mu \cdot \boldsymbol{v}_\mu} \tag{12}$$

The pseudo-temperatures $\Delta T_N$, $\Delta T_R$, and the collective drive velocity $\boldsymbol{u}$ in Eq. (8) are the unknowns to be determined. To close the problem, the momentum- and energy-conserving constraints need to be considered. In hydrodynamic phonon transport, both the R- and N-scattering events conserve the thermal energy:

$$\sum_\mu \frac{1}{\tau_\mu^R}\left(C_\mu \Delta \hat{T}_R - \hat{g}_\mu\right) = 0 \tag{13}$$

$$\sum_\mu \frac{1}{\tau_\mu^N}\left(C_\mu \Delta \hat{T}_N - \hat{g}_\mu\right) = 0 \tag{14}$$

where $\sum_\mu$ denotes the summation of all phonon wavevectors and branches in the Brillouin zone. Substituting Eq. (9) into Eqs. (13-14), a set of relations correlating $\Delta T_R$, $\Delta T_N$, and the drift velocity $\boldsymbol{u}$ is established:

$$\left[\sum_\mu \frac{C_\mu}{\tau_\mu^R}(1-\rho_\mu\chi_\mu)\right]\Delta\hat{T}_R - \left(\sum_\mu \frac{C_\mu}{\tau_\mu^R}\eta_\mu\chi_\mu\right)\Delta\hat{T}_N - \left(\sum_\mu \frac{C_\mu T_0 \boldsymbol{q}}{\omega_\mu \tau_\mu^N}\rho_\mu\chi_\mu\right)\cdot\boldsymbol{u} = \sum_\mu \rho_\mu\chi_\mu p_\mu \tag{15}$$

$$\left(-\sum_\mu \frac{C_\mu}{\tau_\mu^N}\rho_\mu\chi_\mu\right)\Delta\hat{T}_R + \left[\sum_\mu \frac{C_\mu}{\tau_\mu^N}(1-\eta_\mu\chi_\mu)\right]\Delta\hat{T}_N - \left[\sum_\mu \frac{C_\mu T_0 \boldsymbol{q}}{\omega_\mu \tau_\mu^N}(1-\eta_\mu\chi_\mu)\right]\cdot\boldsymbol{u} = \sum_\mu \eta_\mu\chi_\mu p_\mu \tag{16}$$

An additional equation is needed to determine the drift velocity $\boldsymbol{u}$. Since N scattering conserves momentum, the following relation holds:

$$\sum_\mu \frac{\hbar \boldsymbol{q}}{\tau_\mu^N}\left[n_\mu - n_\mu^d(\hat{T}_N)\right] = 0 \tag{17}$$

Substituting Eq. (5) and (9) into Eq. (17), we obtained:

$$-\left(\sum_\mu \frac{C_\mu \boldsymbol{q}}{\omega_\mu \tau_\mu^N}\rho_\mu\chi_\mu\right)\Delta\hat{T}_R + \left[\sum_\mu \frac{C_\mu \boldsymbol{q}}{\omega_\mu \tau_\mu^N}(1-\eta_\mu\chi_\mu)\right]\Delta\hat{T}_N - \left[\sum_\mu \frac{C_\mu T_0 \boldsymbol{q}\boldsymbol{q}^{\mathrm{T}}}{\omega_\mu^2 \tau_\mu^N}(1-\eta_\mu\chi_\mu)\right]\cdot\boldsymbol{u}$$
$$=\sum_\mu \eta_\mu\chi_\mu \frac{p_\mu \boldsymbol{q}}{\omega_\mu} \tag{18}$$

Eqs. (15), (16), and (18) are all equations necessary to solve the unknown pseudo-temperatures and drift velocities, and they can be rewritten in the linear form:

$$\boldsymbol{AX} = \boldsymbol{b} \tag{19}$$

The linear matrix $\boldsymbol{A}$ takes the form:

$$\boldsymbol{A} = \begin{bmatrix} \sum_\mu \frac{C_\mu}{\tau_\mu^R}(1-\rho_\mu\chi_\mu) & -\sum_\mu C_\mu \frac{\tau_\mu}{\tau_\mu^R \tau_\mu^N}\chi_\mu & -\sum_\mu \frac{C_\mu T_0 \boldsymbol{q}^{\mathrm{T}}}{\omega_\mu} \frac{\tau_\mu}{\tau_\mu^R \tau_\mu^N}\chi_\mu \\ -\sum_\mu C_\mu \frac{\tau_\mu}{\tau_\mu^R \tau_\mu^N}\chi_\mu & \sum_\mu \frac{C_\mu}{\tau_\mu^N}(1-\eta_\mu\chi_\mu) & \sum_\mu \frac{C_\mu T_0 \boldsymbol{q}^{\mathrm{T}}}{\omega_\mu \tau_\mu^N}(1-\eta_\mu\chi_\mu) \\ -\sum_\mu \frac{C_\mu \boldsymbol{q}}{\omega_\mu} \frac{\tau_\mu}{\tau_\mu^R \tau_\mu^N}\chi_\mu & \sum_\mu \frac{C_\mu \boldsymbol{q}}{\omega_\mu \tau_\mu^N}(1-\eta_\mu\chi_\mu) & \sum_\mu \frac{C_\mu T_0 \boldsymbol{q}\boldsymbol{q}}{\omega_\mu^2 \tau_\mu^N}(1-\eta_\mu\chi_\mu) \end{bmatrix}, \tag{20}$$

where the wave vector $\boldsymbol{q}$ is written in column, and the superscript T denotes the matrix transpose, $\boldsymbol{q}\boldsymbol{q}$ is the dyadic product. The vector $\boldsymbol{X}$ is the thermal responses in pseudo-temperatures and drift velocities to pulsed point heat source, and the vector $\boldsymbol{b}$ collects the energy and momentum generation rates:

$$\boldsymbol{X} = \begin{bmatrix} \Delta\hat{T}_R \\ \Delta\hat{T}_N \\ \boldsymbol{u} \end{bmatrix}, \qquad \boldsymbol{b} = \begin{bmatrix} \sum_\mu \rho_\mu\chi_\mu p_\mu \\ \sum_\mu \eta_\mu\chi_\mu p_\mu \\ \sum_\mu \eta_\mu\chi_\mu p_\mu \boldsymbol{q}/\omega_\mu \end{bmatrix} \tag{21}$$

The unknown response vector can be easily solved by $\boldsymbol{X} = \boldsymbol{A}^{-1}\boldsymbol{b}$ at each transformation coordinate $(\Omega, \boldsymbol{\xi})$.

Nevertheless, components of $\boldsymbol{X}$ (pseudo-temperature changes and drift velocity) do not directly correspond to experimental observables. The measurable local temperature rise $\Delta T = T - T_0$ is defined as the magnitude of total energy deviation from the equilibrium. We therefore consider the total energy conservation:

$$\sum_\mu \frac{1}{\tau_\mu}\left(C_\mu \Delta\hat{T} - \hat{g}_\mu\right) = 0 \tag{22}$$

By adding Eqs. (13-14), we can obtain:

$$\sum_\mu \left(\frac{C_\mu}{\tau_\mu^R}\Delta\hat{T}_R + \frac{C_\mu}{\tau_\mu^N}\Delta\hat{T}_N\right) = \sum_\mu \frac{\hat{g}_\mu}{\tau_\mu} \tag{23}$$

With Eqs (22-23), the local temperature rise is expressed as:

$$\Delta\hat{T} = \frac{1}{\sum_\mu C_\mu \tau_\mu^{-1}} \sum_\mu C_\mu \left(\frac{\Delta\hat{T}_R}{\tau_\mu^R} + \frac{\Delta\hat{T}_N}{\tau_\mu^N}\right) \tag{24}$$

Clearly, $\Delta\hat{T}$ is a combined result of the R-scattering and N-scattering processes. Therefore, once we solve the response vector $X$, we can express the Green's function of temperature rise as:

$$\hat{\mathcal{G}}_{\Delta T}(\Omega, \boldsymbol{\xi}) = \frac{1}{\sum_\mu C_\mu \tau_\mu^{-1}} \sum_\mu C_\mu \left(\frac{X_R}{\tau_\mu^R} + \frac{X_N}{\tau_\mu^N}\right) \tag{25}$$

where $X_R$ and $X_N$ denote the first two components of $X$ which are the pseudo-temperature changes $\Delta T_R$ and $\Delta T_N$ in response to a unit point pulse.

As illustrated in Figure 1, the temperature response is a combined result of the momentum-conserving N processes and the momentum-destroying R processes. After the $\mathcal{G}_{\Delta T}$ is obtained, temperature response to an arbitrary any heating profile $P(t, \boldsymbol{r})$ can be calculated as the convolution $\Delta T(t, \boldsymbol{r}) = \mathcal{G}_{\Delta T}(t, \boldsymbol{r}) * P(t, \boldsymbol{r})$ in the real domain. In the Fourier domain, the temperature rise is simply the product of the Green's function and the heating profile:

$$\Delta\hat{T}(\Omega, \boldsymbol{\xi}) = \hat{\mathcal{G}}_{\Delta T}(\Omega, \boldsymbol{\xi})\hat{P}(\Omega, \boldsymbol{\xi}) \tag{26}$$

For pump-probe experiments, the thermal signal is measured using a certain sensing profile $S(\boldsymbol{r})$. The measured temperature change can therefore be calculated by averaging $\Delta\hat{T}(\Omega, \boldsymbol{\xi})$ over the Fourier-domain sensing profile $\hat{S}(\boldsymbol{\xi})$:

$$\Delta\hat{T}_S(\Omega) = \int \Delta\hat{T}(\Omega, \boldsymbol{\xi})\hat{S}(\boldsymbol{\xi})d\boldsymbol{\xi} \Big/ \int \hat{S}(\boldsymbol{\xi})d\boldsymbol{\xi} \tag{27}$$

The measured temperature change $\Delta \bar{T}_S(t)$ in time-domain can be simply obtained by inverse Fourier transform $\Delta \bar{T}_S(t) = \int \Delta \hat{\bar{T}}_S(\Omega) e^{i\Omega t} d\Omega/2\pi$.

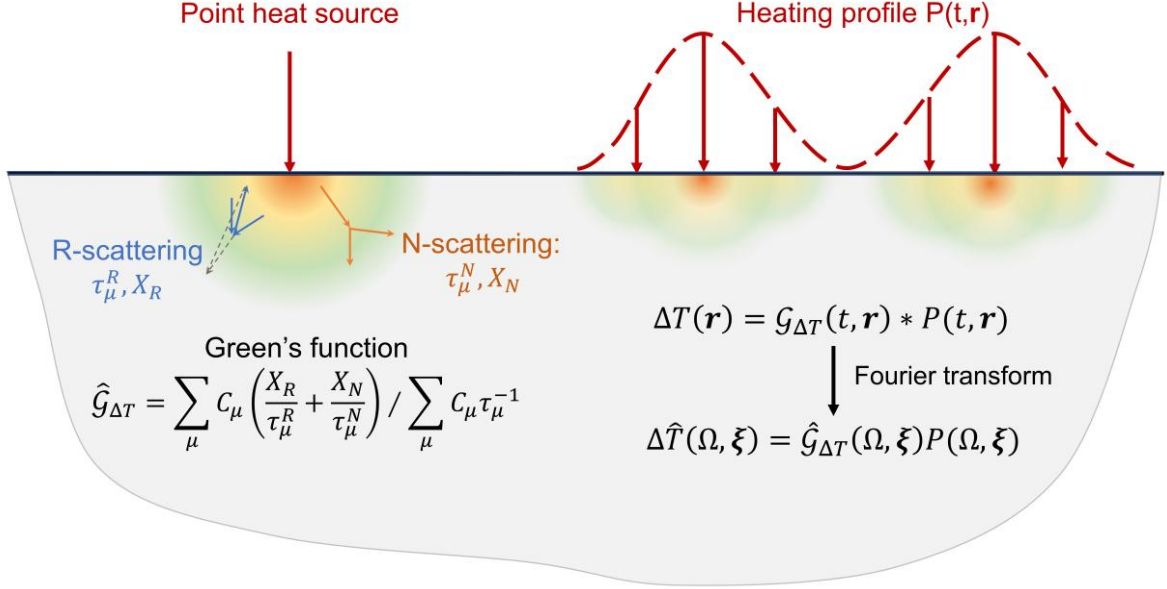

**Figure 1. Illustration of the GF-CSA.** The Green's function is a combined result of the momentum-conserving N-processes and momentum-destroying R-processes. For any heating profile, the temperature response can be calculated by performing a convolution of Green's function with the heating profile in real space, which corresponds to a simple multiplication in the Fourier transformation domain.

While the GF-CSA formalism presented here is limited to semi-infinite domains, this formalism still provides an efficient method to predict heat transport from arbitrary shapes of heat sources using Eqs. (26-27). As we shall see in later sections, GF-CSA can be applied to model ultrafast thermal responses to the spatially periodic TTG heating profile and a ring-shaped heat source as two simple examples. Our analytical method is also considerably faster than numerical methods when simulating transient phonon transport in semi-infinite domains. For example, solving BTE using the discrete ordinate method (DOM) typically has the complexity of $\mathcal{O}(N_\mu N_r N_\theta N_t)$ [39], where $N_\mu$ is the total number of phonon states; $N_r$, $N_\theta$, and $N_t$ are the number of descrte spatial coordinates, number of discrete propagation angles of phonons, and the number of timesteps, respectively. The Monte Carlo method for solving BTE scales similarly as DOM, but it typically involves performing several indendent runs to obtain the ensemble average[40]. In our formalism, the full BTE is reduced to a linear system $\boldsymbol{AX} = \boldsymbol{b}$ for each Fourier domain coordinate $(\Omega, \boldsymbol{\xi})$. The propagation directions have been intrinsically taken care of using the *ab initio* phonon properties by summing over all wave vectors and branches. The computational complexity for evaluating $\boldsymbol{A}$ and $\boldsymbol{b}$ is $\mathcal{O}(N_\mu N_r N_t)$. Since the linear coefficient matrix $\boldsymbol{A}$ at each $(\Omega, \boldsymbol{\xi})$ is simply a 6×6 matrix, the complexity of solving Eq. (19) is further reduced to $\mathcal{O}(N_r N_t)$. As a result, computing signals of TTG and ring-TTR experiments in general take only 20 seconds and a few minutes respectively, even with the script language Python and no parallelization of the program. In comparison, implementing DOM and MC methods typically requires parallelization or GPU programming [41, 42]. Another advantage of the GF-CSA is that, once Green's function is calculated, one can save the data of Green's function and use it to compute phonon transport from different geometries of heat sources simply using Eq. (26) or Eq. (27) without solving the BTE again.

### III. ASYMPTOTIC BEHAVIORS OF GF-CSA

We now examine the asymptotic behaviors of the GF-CSA. In Section III.A, we first show the emergence of second sound waves when N-scattering dominates the phonon-phonon interaction by perturbatively expanding the susceptibility $\chi_\mu$. In Section III.B, we show the recovery to RTA at the strong R-scattering limit, and the convergence to the diffusive transport when $\Omega \tau_\mu \ll 1$ and the ballistic limit when $\Omega \tau_\mu \gg 1$, respectively.

#### A. Emergence of Second Sound with Dominating N-scattering

We analytically show that the damped wave equation of the second sound emerges by perturbatively expanding the susceptibility $\chi_\mu$ at the limit $\tau_\mu^N \ll \tau_\mu^R$, *i.e.* N-scattering dominates the phonon-phonon interaction. For mathematical simplicity, we consider the homogenous condition with the source term $b = 0$. Without losing the general physics, one-dimensional transport is assumed here to show the asymptotic damped wave behavior of the GF-CSA. When N processes dominate over R processes, $\tau_\mu \approx \tau_\mu^N \ll \tau_\mu^R$, then $\Delta T \approx \Delta T_N$ and $\Delta T_R$ becomes negligible according to Eq. (25). The dimensionality of the linear equation $AX = b$ is therefore reduced to:

$$\begin{bmatrix} \sum_\mu \frac{C_\mu}{\tau_\mu^N}(1-\eta_\mu \chi_\mu) & \sum_\mu \frac{C_\mu T_0 q_x}{\omega_\mu \tau_\mu^N}(1-\eta_\mu \chi_\mu) \\ \sum_\mu \frac{C_\mu q_x}{\omega_\mu \tau_\mu^N}(1-\eta_\mu \chi_\mu) & \sum_\mu \frac{C_\mu T_0 q_x^2}{\omega_\mu^2 \tau_\mu^N}(1-\eta_\mu \chi_\mu) \end{bmatrix} \begin{bmatrix} \Delta T \\ u \end{bmatrix} = \mathbf{0} \tag{28}$$

In the limit $F_{\mu,x}\xi_x \ll 1$ and $\Omega \tau_\mu \ll 1$, the susceptibility can be expanded as:

$$1 - \eta_\mu \chi_\mu = \rho_\mu + \eta_\mu(i\Omega\tau_\mu + iF_{\mu,x}\xi_x) \tag{29}$$

To the linear order, the matrix elements can be expanded as:

$$\sum_\mu \frac{C_\mu}{\tau_\mu^N}(1-\eta_\mu \chi_\mu) \approx i\Omega C \tag{30}$$

$$\sum_\mu \frac{C_\mu T_0 q_x}{\omega_\mu \tau_\mu^N}(1-\eta_\mu \chi_\mu) \approx \sum_\mu \frac{C_\mu T_0 q_x}{\omega_\mu} v_{\mu,x} \cdot i\xi_x \tag{31}$$

$$\sum_\mu \frac{C_\mu q_x}{\omega_\mu \tau_\mu^N}(1-\eta_\mu \chi_\mu) \approx \sum_\mu \frac{C_\mu q_x}{\omega_\mu} v_{\mu,x} \cdot i\xi_x \tag{32}$$

$$\sum_\mu \frac{C_\mu T_0 q_x^2}{\omega_\mu^2 \tau_\mu^N}(1-\eta_\mu \chi_\mu) \approx \sum_\mu \frac{C_\mu T_0 q_x^2}{\omega_\mu^2}\left(\frac{1}{\tau_\mu^R} + i\Omega\right) \tag{33}$$

With inverse Fourier transform, $i\Omega$ and $i\xi_x$ are replaced with differential operators $\frac{\partial}{\partial t}$ and $\frac{\partial}{\partial x}$, and a set of partial differential equations can be derived:

$$C\frac{\partial T}{\partial t} + \left(\sum_\mu \frac{C_\mu T_0 q_x v_{\mu,x}}{\omega_\mu}\right)\frac{\partial u}{\partial x} = 0 \tag{34}$$

$$\left(\sum_\mu \frac{C_\mu q_x v_{\mu,x}}{\omega_\mu}\right)\frac{\partial T}{\partial x} + \left(\sum_\mu \frac{C_\mu T_0 q_x^2}{\omega_\mu^2}\right)\frac{\partial u}{\partial t} = -\left(\sum_\mu \frac{C_\mu T_0 q_x^2}{\omega_\mu^2 \tau_\mu^R}\right)u \tag{35}$$

Combining these two equations, the damped-wave equation for both temperature and the drift velocity $u$ can be derived:

$$\frac{\partial^2 T}{\partial t^2} + \frac{1}{\tau_{ss,x}} \frac{\partial T}{\partial t} - v_{ss,x}^2 \frac{\partial^2 T}{\partial x^2} = 0 \tag{36}$$

$$\frac{\partial^2 u}{\partial t^2} + \frac{1}{\tau_{ss,x}} \frac{\partial u}{\partial t} - v_{ss,x}^2 \frac{\partial^2 u}{\partial x^2} = 0 \tag{37}$$

$\tau_{ss,x}$ and $v_{ss,x}^2$ are relaxation time and second sound velocity along the $x$-direction:

$$\frac{1}{\tau_{ss,x}} = \frac{\sum_\mu C_\mu q_x^2 / (\omega_\mu^2 \tau_\mu^R)}{\sum_\mu C_\mu q_x^2 / \omega_\mu^2}, \tag{38}$$

$$v_{ss,x}^2 = \frac{\left(\sum_\mu C_\mu q_x v_{\mu,x} / \omega_\mu\right)^2}{C \sum_\mu C_\mu q_x^2 / \omega_\mu^2}. \tag{39}$$

Eqs. (36-37) show that both the temperature field and the drift velocity behave as attenuating waves, and the damping rate is determined by the R-scattering strength $1/\tau_\mu^R$. Results obtained by perturbative expansion of GF-CSA are consistent with previous derivations by Hardy [24] and Ding et al [17].

### B. Convergence to RTA, Diffusive, and Ballistic Limits

The GF-CSA developed in this work reduces the RTA when R-scattering dominates over N-scatterings, and further recovers the Fourier solution at the diffusive limit and the ballistic solution with $\Omega \tau_\mu \gg 1$. With dominating R-scattering, $\tau_\mu \approx \tau_\mu^R \ll \tau_\mu^N$, only the matrix element $A_{11}$ approaches to $\sum_\mu C_\mu \tau_\mu^{-1}(1 - \chi_\mu)$, and the other matrix elements vanishes. The source term $\boldsymbol{b} \to \left[\sum_\mu \chi_\mu p_\mu, 0, 0\right]^T$. As a result, Green's function of temperature change at the R-scattering limit (i.e. the RTA) is simply solved as $b_1/A_{11}$:

$$\hat{\mathcal{G}}_{\Delta T}^{RTA}(\Omega, \boldsymbol{\xi}) = \frac{\sum_\mu \chi_\mu p_\mu}{\sum_\mu \frac{C_\mu}{\tau_\mu}(1 - \chi_\mu)} \tag{40}$$

This is identical to the analytical results derived by Hua et al [29]. At the limit $\Omega \tau_\mu \ll 1$ and $|\boldsymbol{\Lambda}_\mu \cdot \boldsymbol{\xi}| \ll 1$, leading order expansion of the susceptibility is $\chi_\mu \approx 1 + i\Omega \tau_\mu + (\boldsymbol{F}_\mu \cdot \boldsymbol{\xi})^2$:

$$\hat{\mathcal{G}}_{\Delta T}^{RTA} \approx \frac{1}{\sum_\mu \frac{C_\mu}{\tau_\mu}\left[i\Omega\tau_\mu + (\boldsymbol{F}_\mu \cdot \boldsymbol{\xi})^2\right]} = \frac{1}{i\Omega C + \sum_j K_j \xi_j^2} = \hat{\mathcal{G}}_{\Delta T}^F \tag{41}$$

where $K_j = \sum_\mu C_\mu F_{\mu,j}^2 / \tau_\mu = \sum_\mu C_\mu v_{\mu,j} F_{\mu,j}$ is the thermal conductivity along the $j$-th direction, and $\hat{\mathcal{G}}_{\Delta T}^F$ denotes the Green's function of the heat conduction equation using Fourier's law.

Finally, the ballistic regime occurs when $\Omega \tau_\mu \gg 1$, and the BTE is reduced to:

$$\left(\frac{\partial}{\partial t} + \boldsymbol{v}_\mu \cdot \nabla\right) g_\mu = p_\mu \delta(\boldsymbol{r})\delta(t) \tag{42}$$

The modal energy deviations can be solved by performing Fourier transforms, and ballistic temperature profile [29] is recovered:

$$\hat{\mathcal{G}}_{\Delta T}^b(\boldsymbol{\xi}, \Omega) = \frac{1}{C} \sum_\mu \frac{p_\mu}{i[\Omega + \boldsymbol{v}_\mu \cdot \boldsymbol{\xi}]} \tag{43}$$

It is clear that Green's function $\hat{\mathcal{G}}_{\Delta T}^b$ at the ballistic limit shows a phase lag of $\pi/2$ to the input heating, due to the imaginary unit $i$ in the denominator of Eq. (43). This is quite different when hydrodynamic second sound is observed, which allows a phase inversion by $\pi$ due to the $\partial^2 T/\partial t^2$ term in Eq. (36). As we shall see in Section IV, the $\pi$ phase shift is manifested as a negative temperature dip after a heat pulse in pump-probe experiments, due to the constructive interference of second sound waves.

**IV. MODELING TRANSIENT RESPONSE IN PUMP-PROBE MEASUREMENTS**

In this section, we apply the GF-CSA formalism for modeling the ultrafast thermal response under pulsed heating in pump-probe measurements. Ultrafast temperature oscillations are analyzed in the in-plane direction of graphite, in transient thermal grating (TTG) and transient thermoreflectance (TTR). In the former case, the thermal transport takes place across the periodic grating, and the second sound corresponds to one-dimensional plane waves. In TTR, the thermal transport has radial symmetry if the probe beam and the pump beam are co-axial, and the second sound corresponds to cylindrical waves. Such radial symmetry can be broken once the probe beam has a finite offset. The GF-CSA formalism can be conveniently incorporated to model the ultrafast temperature oscillations, and successfully captures the signature of the second sound in the aforementioned geometries, manifested as the temperature dipping below the initial temperature at a few nanoseconds. To obtain the phonon properties of graphite, we performed first-principles calculations using the Quantum Espresso package [43]. Harmonic force constants are calculated using the density functional perturbation theory (DFPT) using the Perdew-Zunger (PZ) exchange-correlation functionals[44] with a cutoff energy of 80 Rydberg and a 10×10×4 supercell. The third-order force constants are calculated using the finite displacement method with a 5×5×4 supercell within a cutoff of 4 Å, using the thirdorder.py script of ShengBTE package[25] with a 32×32×10 mesh in the momentum space. Phonon lifetimes, and mean free displacements are obtained by iteratively solving the BTE using the ShengBTE package [25].

*One-dimensional periodic heating in TTG.* We first model the simplest one-dimensional hydrodynamic phonon transport in TTG. As illustrated in Figure 2a, TTG uses two pulsed laser beams cross-focused at the sample surface to create an interference pattern, and the heat source imposed on the sample has a spatially periodic profile $P(t,x) = P_0 \cos(2\pi x/L)\delta(t)$ with $L$ the grating period and $P_0$ the pump intensity, as shown in Figure 2b. Such sinusoidal heating creates a rippled surface expansion that acts as a transient diffraction grating for the probe laser light. The diffracted probe light is directed to the photodetector for measuring the surface temperature change. Such spatially periodic heating profile corresponds to two spikes in the Fourier domain at $\xi_x = \pm 2\pi/L$, *i.e.* $\hat{P}(\Omega, \xi_x) = P_0\left[\delta(\xi_x + 2\pi/L) + \delta(\xi_x - 2\pi/L)\right]/2$. The probe light is a CW laser, whose radius is much larger than the grating period, thus can be approximately regarded as a uniform profile $S(t, \boldsymbol{r}) = S_0$. Using Eq. (27), the $\Delta\hat{\bar{T}}_S(\Omega)$ are simply adding Green's functions at $\pm 2\pi/L$, and time-domain changes $\Delta\bar{T}_S(t)$ can be calculated using the inverse Fourier transform:

$$\Delta \bar{T}_S(t) = \frac{P_0}{4\pi} \int_{-\infty}^{\infty} \left[ \mathcal{G}_{\Delta T}\left(\Omega, \frac{2\pi}{L}\right) + \mathcal{G}_{\Delta T}\left(\Omega, -\frac{2\pi}{L}\right) \right] e^{i\Omega t} d\Omega \tag{44}$$

Figure 2c shows the evolution of normalized temperature rise $\Delta \bar{T}_S(t) / \max(\Delta \bar{T}_S)$ from 50 K to 300 K, with the grating size $L = 3$ μm. Near room temperatures (250 K to 200 K), $\Delta \bar{T}_S$ shows a monotonic decay with increasing time, indicating a diffusive behavior. Between 80 K and 150 K, $\Delta \bar{T}_S$ shows oscillations with a negative below zero. Such a negative dip is attributed to a phase shift by $\pi$: temperatures in the region heated by the thermal grating become even cooler than the initial temperature. Such negative dips indicate the temperature field behaves like a standing wave, which is the signature of the hydrodynamic second sound. At a very low temperature of 50 K, the phonon transport is primarily ballistic hence no lattice cooling effect is observed [16]. By scanning over different temperatures and grating periods, the hydrodynamic window can be mapped as shown in Figure 2d. The second sound is observed when the characteristic size $L$ of the temperature field is small enough such that phonons are dissipated before experiencing momentum-destroying R-scattering events, but still large enough to allow enough N-scattering. Namely, hydrodynamic lattice cooling due to second sound only existed when $1/\tau_\mu^R < v_\mu/L < 1/\tau_\mu^N$. In Figure 2e-f, we compared the temperature evolution predicted by GF-CSA with the TTG experimental data, showing reasonable agreements in both the dipping time and the dipping depths. Based on the dipping time, the apparent second sound speed is calculated as $v_{ss} = L/(2t_{dip}) = 3.9$ km/s at 100 K and 3.5 km/s at 150 K, agreeing well with previous reports [16, 17]. Such propagation speed is much slower than the group velocity of LA (~21 km/s) and TA (~12 km/s) phonons in graphite, ruling out the possibility of temperature oscillations due to ballistic phonon transport. The decreased $v_{ss}$ with increasing temperatures is also consistent with experimental observation[17], due to the larger fraction of diffusive phonons caused by frequent phonon-phonon scattering at increased temperatures. Compared with TTG responses calculated with the full scattering matrix (FSM), our formalism successfully reproduced the hydrodynamic dipping time. The predicted magnitude of the dip is different from the FSM predictions [17], but still shows reasonable agreement with the experiment. Some fluctuation features in Figure 2e in experimental TTG signals are not captured by our model using GF-CSA. There are two possible reasons for such discrepancy. First, phonon properties calculated at finite Brillouzin zone mesh have imposed an artificial cutoff on long wavelengths of phonons, which can propagate ballistically for long distances and contribute to extra fluctuations upon the TTG signal. Second, the GF-CSA represented a physically reasonable simplification of the full scattering matrix. As a result, some selection rules for phonon-phonon interactions are not directly captured in the GF-CSA formalism, and our formalism might tend to suppress ballistic fluctuations.

Figure 3 shows the difference in temperature oscillations in the ballistic limit and with hydrodynamic second sound. Figure 3a shows the TTG response $\Delta \bar{T}_S$ in the time domain with a very small grating $L = 0.5$ μm where phonons transport ballistically. In this case, our formalism using CSA approaches RTA solution, and the temperature oscillations show no pronounced negative dips. In the frequency-domain representation (Figure 3b), the temperature response $\Delta \hat{\bar{T}}_S(\Omega)$ decrease monotonically with the increasing frequency $\Omega$. However, when the hydrodynamic second sound occurs ($L = 4$ μm), the negative temperature dip becomes pronounced (Figure 3c), and the frequency-domain response $\Delta \hat{\bar{T}}_S(\Omega)$ shows a broadened resonant peak (Figure 3d). Such resonant peak also manifests the existence of the second sound wave [16], which is absent when RTA or Fourier theories are used to model the TTG responses.

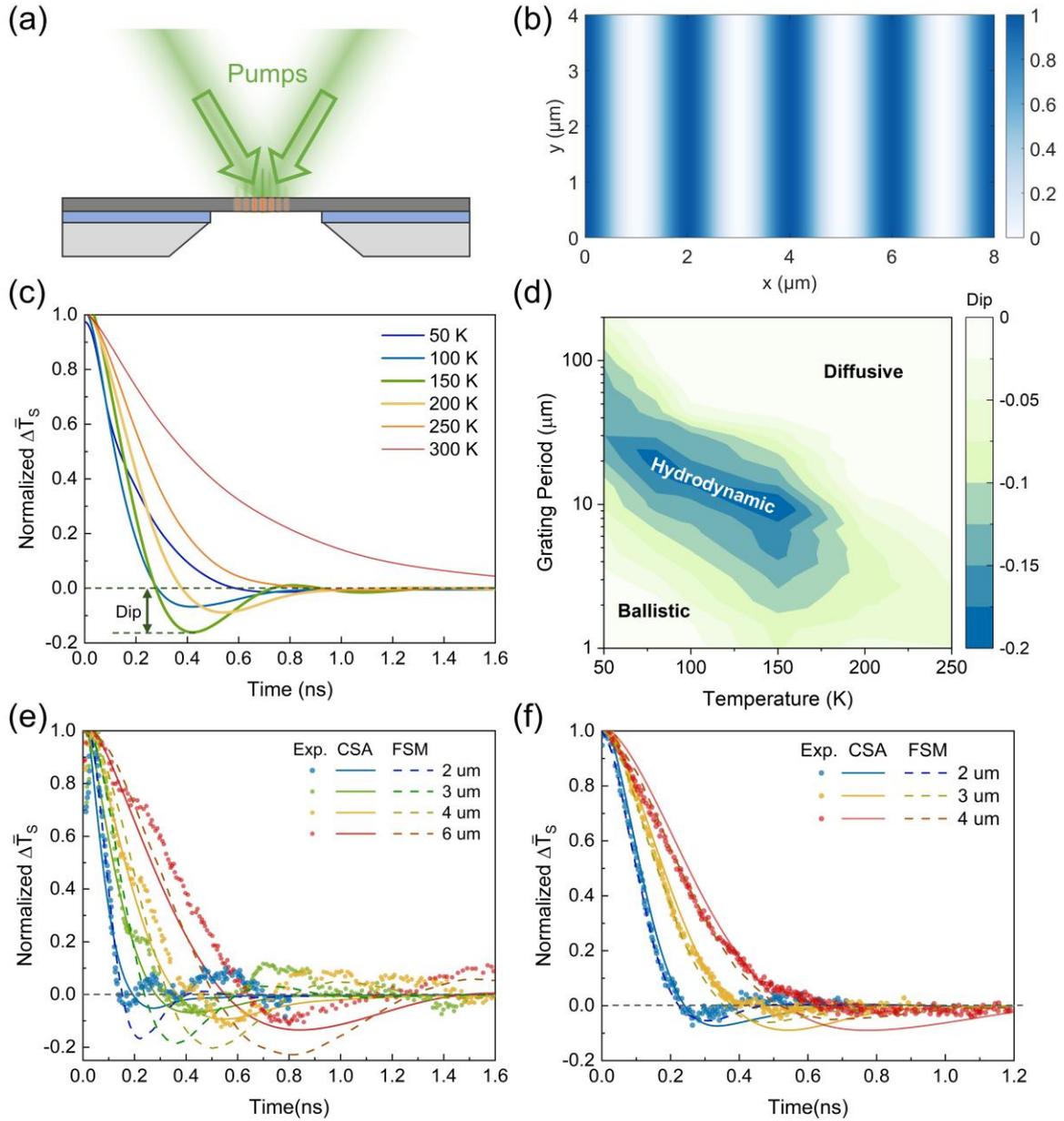

**Figure 2. Temperature response in graphite under transient thermal grating (TTG)**. (a) Illustration of the TTG heating scheme. (b) Periodic heating intensity with the color bar showing the minimum heating rate to the maximum heating. (c) Temperature-dependent thermal responses at the grating period $L = 3$ μm. The temperature dip below zero indicates the lattice cooling due to the hydrodynamic second sound waves. (d) The predicted hydrodynamic window using GF-CSA. (e-f) Grating-period-dependent temperature responses at (e) 100 K, and (f) 200 K. Solid lines are our results using GF-CSA. Experimental data (symbols) and calculated signal with full scattering matrix (FSM, dashed lines) are from ref. 17. All $\Delta \bar{T}_S$ are normalized by the maximum temperature rise.

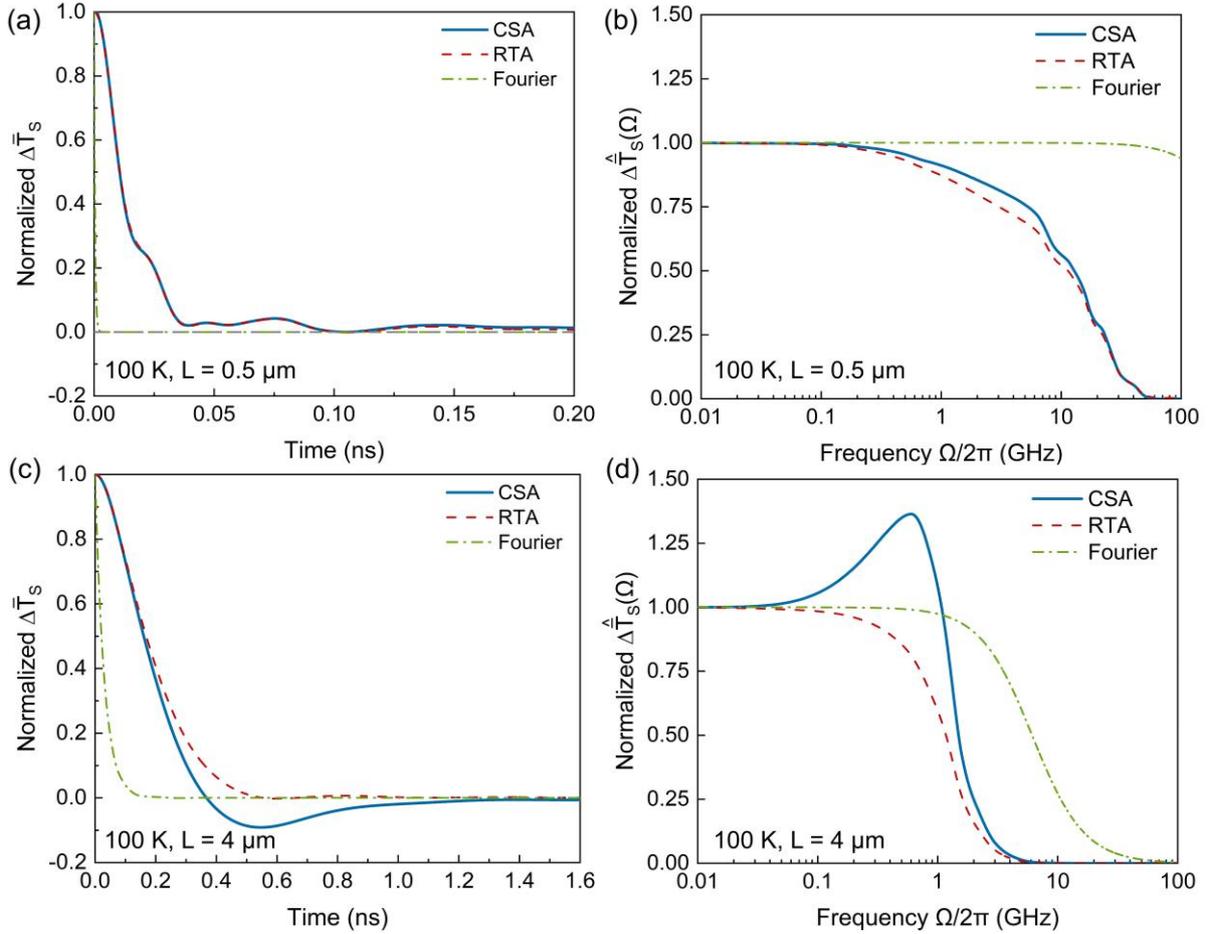

**Figure 3. Ballistic and Hydrodynamic temperature oscillations in the time domain and frequency domain.** (a) The temperature oscillations in $\Delta \bar{T}_S$ at the ballistic limit, with $T_0 = 100$ K and a grating period of 0.5 μm. (b) The frequency-domain representation $\Delta \hat{\bar{T}}_S(\Omega)$ shown in panel (a). (c) The temperature oscillations in $\Delta \bar{T}_S$ in the hydrodynamic window, with $T_0 = 100$ K and a grating period of 4 μm. (d) The frequency-domain representation $\Delta \hat{\bar{T}}_S(\Omega)$ shown in panel (c). $\Delta \hat{\bar{T}}_S(\Omega)$ are normalized by the low-frequency limit $\Delta \hat{\bar{T}}_S(\Omega = 0)$.

*Radially symmetric Ring-TTR.* In addition to the simple TTG, we have further modeled the transient thermoreflectance using a ring-shaped pump beam (Ring-TTR), as shown in Figure 4a. Different from conventional transient thermoreflectance using a normal Gaussian pump beam, the Ring-TTR uses a ring-shaped pump that excites thermal phonons at the ring and allows the phonons to propagate a certain distance to the ring center. Due to the radial symmetry, the thermal waves reaching the center are in-phase, and will constructively interfere to amplify the temperature oscillations due to phonon hydrodynamics. The temperature oscillations would result in reflectance change proportional to $\Delta T$, which is then measured by a CW probe beam.

The probe light reflected from the surface can be directed to photodetectors, thus the measured signal is proportional to the temperature rise $\Delta \bar{T}_S$ weighted by the probe profile. In the real domain, the profiles of the pulsed ring-pump ($P$) and CW probe ($S$) beams are:

$$P(t, \boldsymbol{r}) = \frac{P_0}{\pi w_p^2} \exp\left[-\frac{2(|\boldsymbol{r}| - R_0)^2}{w_p^2}\right] \delta(t), \quad (45)$$

$$S(\boldsymbol{r}) = \frac{S_0}{\pi w_s^2} \exp\left[-\frac{2|\boldsymbol{r}|^2}{w_s^2}\right] \quad (46)$$

where $P_0$ and $S_0$ are the intensities of the pump and the probe beams respectively; $R_0$ is the radius of the ring; $w_p$ is the $1/e^2$ half-width of the ring, and $w_s$ is the $1/e^2$ radius of the probe. The spatial part of the pump profile can be regarded as the convolution of a Gaussian profile and a radial delta function: $\exp[-2(|\boldsymbol{r}| - R_0)^2/w_p^2] = \exp[-2|\boldsymbol{r}|^2/w_p^2] * \delta(|\boldsymbol{r}| - R_0)$, and its Fourier transform is therefore simply the product of the corresponding Fourier-domain representations. The Fourier-transformed profiles of the pump ($\hat{P}$) and the probe ($\hat{S}$) beams are therefore written as:

$$\hat{P}(\boldsymbol{\xi}) = P_0 \exp\left[-\frac{w_p^2 \xi^2}{8}\right] \cdot J_0(|\boldsymbol{\xi}| R_0), \quad (47)$$

$$\hat{S}(\boldsymbol{\xi}) = S_0 \exp\left[-\frac{w_s^2 \xi^2}{8}\right] \quad (48)$$

Here, $\boldsymbol{\xi} = (\xi_x, \xi_y)$ is the two-dimensional Fourier transform vector, and $J_0$ is the 0-th order Bessel function. The measured temperature response can then be calculated using Eq. (27) in the frequency domain, and the time-domain results are further obtained using the inverse Fourier transform. With co-axial pump and probe beams (Figure 4b), the predicted temperature responses from 80 K to 200 K are shown in Figure 4c. Our model captured the transient lattice cooling at 80 K and 100 K, with a dip located at 6 ns agreeing well with experiments. Theoretical calculations, either using our GF-CSA formalism or solving BTE with the full scattering matrix with Monte Carlo (FSM-MC), show a much weaker hydrodynamic dip than the experiments. The GF-CSA also predicts larger times at maximum temperature rise than FSM-MC and experiments. This is because our formalism deals with a semi-infinite body, while there are grain boundaries in the graphite sample used for measurements, and adiabatic boundaries are assumed in FSM-MC simulations. Nevertheless, our formalism still captures the hydrodynamic dip at 100 K, while such a dip has already vanished in FSM-MC simulations[20]. Possible reasons for the shallower dip include (1) the limited signal-to-noise ratio in the ring-TTR experiment, with the measured signal even fluctuating below zeros at high temperatures (200 – 300 K) in the experimental data [20]; (2) the limited mesh in the Brillouin zone imposes a cutoff for the phonon wavelengths, such that long-wave-length phonons whose scattering is dominated by N processes are not included in the modeling. Such long-wavelength cutoff might also be responsible for the absence of second sound dips at 150 K in first-principles modeling, while the experimental data still shows a dip touching the zero-temperature rise limit. On the other hand, significant noise in the ring-TTR signal might also introduce uncertainties in the dip depths. The original experimental data at 294 K reported even negative $\Delta \bar{T}_S$ at long delay times (15 ~ 20 ns) [20], where no hydrodynamic second sound should occur. Second sound velocity can also be extracted as $v_{ss} = R_0/\Delta t_{pd}$ using to the peak-to-dip time $\Delta t_{pd}$. Using the calculated ring-TTR signal, the apparent second sound speed is calculated as ≈ 3.8 km/s at 100 K, agreeing well with the intrinsic second sound ~ 3.9 km/s [17].

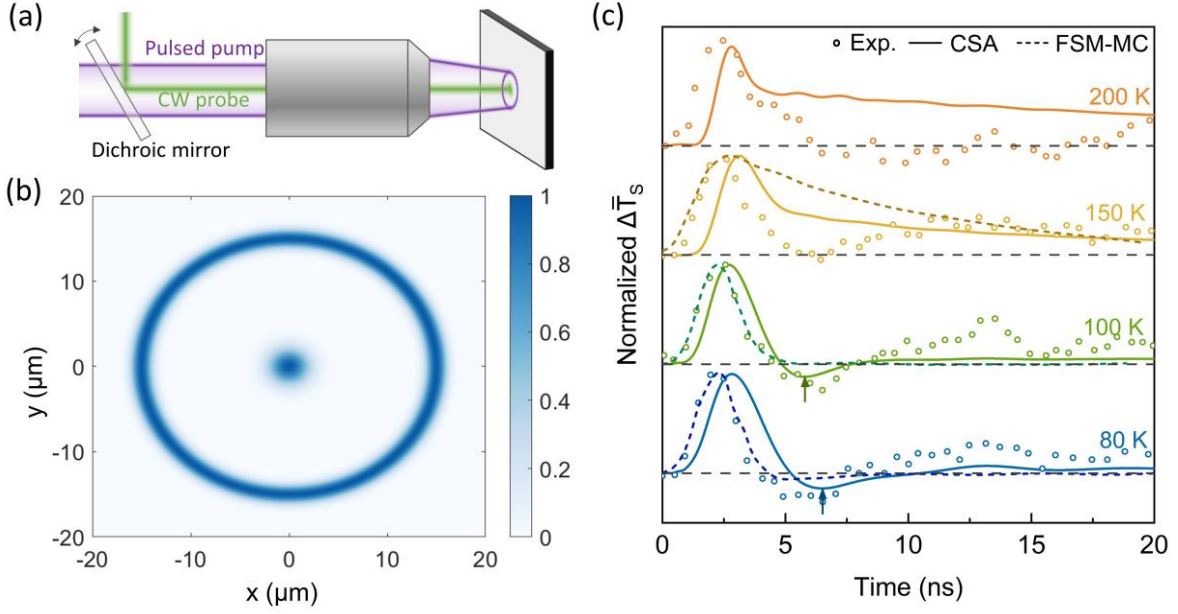

**Figure 4. Temperature response in ring-pumped transient thermoreflectance (ring-TTR).** (a) Illustration of ring-TTR. The sample is heated by a pulsed laser with a ring shape, and the CW probe beam probes the temperature in the ring center. (b) Beam intensities of co-axial ring pump and Gaussian probe, with a pump-ring radius of 15 μm and a width of 3 μm, and a probe beam diameter of 6 μm. (c) Modeled temperature responses in this work (solid lines) compared with experimental data (symbols) and calculated signals by solving the BTE with full scattering matrix using the Monte Carlo method (FSM-MC, dashed lines) [20].

*Off-centered probe in ring-TTR.* After modeling co-axial pump and probe beams with radial symmetry, we further investigated ring-TTR with an off-centered probe breaking the radial symmetry. The TTR signal can be easily calculated by integrating the GF-CSA over the pump and probe profiles in the Fourier domain, with the off-centered probe written as $S(\boldsymbol{\xi}, \boldsymbol{r}_{off}) = S(\boldsymbol{\xi}) \cdot \exp[-i\boldsymbol{\xi} \cdot \boldsymbol{r}_{off}]$. The laser intensities and the predicted TTR signals are shown in Figure 5a-b. As the probe beam moves away from the center towards the pump ring, there is increased overlap between the pump intensity and the probe intensity, and the instantaneous temperature rise is picked up by the probe beam, which results in the increased signal magnitude at $t = 0$. The peak time is also delayed with increasing offset, because it takes a longer time for the thermal excitation from the other side of the ring to reach the probe spot. The probe beam offset imposes a spatial phase on the thermal response, therefore shifting the location of temperature peaks and dips. By scanning the probe within the ring-pumped area, we can also obtain the probe-weighted temperature profile at $t = 2.85$ ns and $t = 6.78$ ns, which corresponds to the maximum temperature rise and lattice cooling with zero probe offset, as shown in Fig. 5c-d. Maximum hydrodynamic lattice cooling is measured when the probe beam is located at the center. With cylindrical heating, the ring center is equivalent to an adiabatic boundary condition, and the heat wave incident to the ring center is reflected with a phase inversion, which leads to maximum lattice cooling near the center.

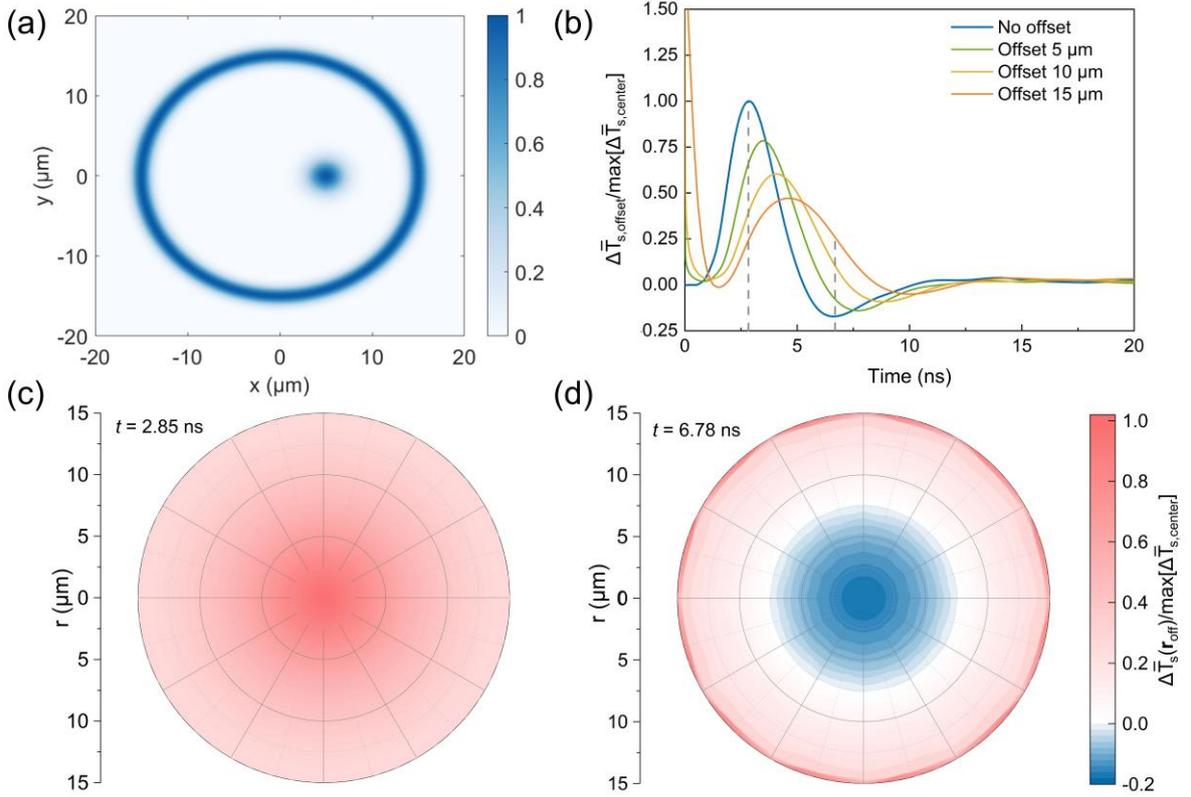

**Figure 5. Off-centered probe measurement.** (a) Off-centered pump and probe beam intensities. (b) Temperature oscillations at different beam offsets. The response is normalized by the maximum temperature rise with the probe beam at the ring center. Temperature profiles at (c) 2.85 ns and (d) 6.78 ns, with the time slices indicated by dashed lines in (b).

## V. SUMMARY

We have derived Green's function of the multidimensional Boltzmann transport equation under Callaway scattering approximation for modeling hydrodynamic second sound. We also showed the emergence of second sound waves by the perturbative expansion of Green's function, and we mathematically proved the convergence to ballistic and diffusive transport at the limits of $\Omega\tau_\mu \gg 1$ and $\Omega\tau_\mu \ll 1$, respectively. This formalism can be conveniently implemented to predict transient temperature responses to arbitrary-shaped heat sources in infinite or semi-infinite domains, using fully first-principle phonon dispersions and lifetimes. The modeling prediction shows nice agreement in temperature oscillation dynamics and second sound velocities with the literature reports in TTG and ring-TTR measurements. Our work provides a simple and efficient formalism for modeling transient hydrodynamic phonon transport in different heating geometries.

**Acknowledgment**

This work is supported by the National Key R&D Program (No. 2022YFA1203100) and the National Natural Science Foundation (No. 52076089 and No.12147122). The authors declare no conflict of interest.